\documentclass[journal]{IEEEtran}
\usepackage{amsmath,amsthm,amsfonts,amssymb,bm} % 数学宏包
\usepackage{graphicx,psfrag} % 图形宏包

\usepackage{algorithm}
\usepackage{algorithmicx}
\usepackage{algpseudocode}

\newcommand{\bs}{\boldsymbol}

\begin{document} % 开始正文

\title{Symbol Detection for Massive MIMO AF Relays Using Approximate Bayesian Inference}
\author{
Haochuan zhang and Qiuyun Zou

\thanks
{
H. Zhang is with Guangdong University of Technology, Guangzhou 510006, China (email: haochuan.zhang@qq.com).
}

\thanks{
Q. Zou is with Beijing University of Posts and Telecommunications,
Beijing 100876, China (email: qiuyun.zou@bupt.edu.cn).
}
}

\maketitle % 生成标题
%---------------------------------章节部分--------------------------%

%%-------公式垂直间距设置---------%
%\abovedisplayshortskip =0pt %短公式
%\abovedisplayskip =8pt      %长公式
%\belowdisplayshortskip=6pt
%\belowdisplayskip=8pt

\begin{abstract}%75-100 words
For massive MIMO AF relays, symbol detection becomes a practical issue when the number of antennas is not large enough, since linear methods are non-optimal and optimal methods are exponentially complex. This paper proposes a new detection algorithm that offers Bayesian-optimal MSE at the cost of $O(n^3)$ complexity per iteration. The algorithm is in essence a hybrid of two methods recently developed for deep learning, with particular optimization for relay. As a hybrid, it inherits from the two a state evolution formulism, where the asymptotic MSE can be precisely predicted through a scalar equivalent model. The algorithm also degenerates easily to many results well-known when single-hop considered.
\end{abstract}
%本文考虑具有完美信道状态信息互易性的两跳中继通信系统中的大规模多输入多输出（MIMO）接收机。为了这个目的，我们对Flether提出的MLVAMP 做出了修正，提出了the proposed algorithm，该算法主要用深度神经网络。


%\hypersetup{CJKbookmarks=true}
%\begin{spacing}{1.2}    %1.2 倍行距

\section{Introduction}

Massive multiple-input multiple-output (MIMO) \cite{Marzetta-TWC10-UnlimitedAnt} is  currently a compelling sub-6 GHz physical-layer technology for future wireless access, including 5G \cite{Larsson-CMag14-Massive}. It offers many desirable benefits, among which are low complexity processing, excellent spectral efficiency, and superior energy efficiency, by using an unlimited number of antennas \cite{Marzetta-TWC10-UnlimitedAnt}. In the technology's practical rolling out, however, only a limited number of antennas (tens to hundreds) was adopted, considering software and hardware limitations. Such an inadequacy in the antenna number poses a critical challenge to symbol detection, a component key to modern digital architecture that recovers/estimates transmitted data from observations corrupted by noise and channel fading. With only a limited number of antennas, simple linear detectors (as advocated by \cite{Marzetta-TWC10-UnlimitedAnt}) are far away from optimal since the channel matrix is hardly orthogonal \cite{Hoydis-JSAC13-AntNumber}. Theoretically optimal or sub-optimal (nonlinear) detectors, such as the maximum likelihood (ML) and sphere decoding, however, require a computational complexity that grows exponentially with the number of antennas. Considering the case of high-order modulation with several hundreds antennas, this complexity quickly soars up to an unacceptably high level.

One possible way out of the dilemma could be using state-of-the-art algorithms from compressed sensing and artificial intelligence to attain a good approximate Bayesian inference. Among these algorithms are approximate message passing (AMP) \cite{Donoho-PNAS09-AMP}, expectation propagation (EP) \cite{Minka-PhD01-EP}  (a.k.a. expectation consistent, EC \cite{Opper-JMLR05-EC} ), and their variants \cite{Rangan-arxiv10-GAMP, Parker-TSP14-BiGAMP1, Parker-TSP14-BiGAMP2, Fletcher-arxiv16-GEC, He-ISIT17-GEC-SR, He-JSTSP18-GEC-SR}. The AMP class uses a quadratic approximation of the loopy belief propagation to derive an efficient implementation of the Bayesian estimation, requiring only linear complexity per iteration. Another striking aspect of the AMP class is that their asymptotic mean squared error (MSE) can be accurately tracked by a simple one-dimensional iteration termed state evolution (SE) \cite{Bayati-TIT11-SE}.
EP, on the other hand, originating from variational inference, can be applied to a class of weighting matrices broader than AMP. It also enjoys a MSE performance better than the AMP class, but at a price of increasing complexity that is roughly on the order of $O(n^3)$, with $n$ being the dimension of matrix inversion. Both classes can offer a decent tradeoff between efficiency and effectiveness. For this reason, they were soon introduced to the communications community and had gained their popularity ever since. To name a few of the examples, the AMP class has been successfully applied to symbol detection of massive MIMO in \cite{Wu-JTSP14-AMP}, to channel estimation in \cite{Wen-TWC15-ChanEstAMP}, and to joint channel-and-data detection in \cite{Wen-TSP16-JCDBiGAMP}. For the EP class, successful applications can be found in \cite{Qi-TWC07-EPCom,Martin-TCOM14-EP,He-JSTSP18-GEC-SR} and so on. These works, however, all relied on the assumption of a single-hop communication. The methods used cannot be extended to cover the more general framework of multi-hop communications.

Multi-hop communications, a.k.a. relaying, is a key enabler for throughput improvement, as well as coverage enhancement \cite{Laneman-TIT04-Relay}. There are actually two types of relays: one is the amplify-and-forward (AF) (a.k.a. non-regenerative) relay, which simply repeats the physical signal received, and the other is the decode-and-forward (DF) (a.k.a. regenerative) relay that decodes the message before sending it out again. Here we consider AF relay exclusively and assume the use of massive MIMO. Although our method can be easily extended to cover three hops and more, we provide details only for a dual-hop case to ease statements. In the multi-hop massive MIMO setting, we found that previous work were mostly focusing on theoretical analysis \cite{Jin-TWC15-MassiveRelay, Zhao-TWC16-Relay}, with very few on the practical issue of symbol detection. A reason for this would lie in the fundamental limitation that conventional factor graph of the AMP and EP classes involves only two blocks, representing a transmitter and a receiver. No relay is involved or allowed. Given the practical importance of symbol detection in modern digital systems, finding an algorithm that is both effective and efficient, has thus become a issue of urgent need, not only to the industry but also to the academia.

Fortunately, an algorithm termed multi-layer vector AMP (ML-VAMP) \cite{Fletcher-arxiv17-ML-VAMP} may offer a solution, which was recently proposed for deep learning. Belonging to the family of approximate Bayesian inference, the algorithm is capable of handling the concatenation of multiple standard linear models (i.e., linear mixing plus Gaussian noise) and multiple nonlinear activation functions. We find that such a concatenation resembles the multi-hop relay in certain aspects, and also that another algorithm, termed multi-layer generalized expectation consistent (ML-GEC) \cite{Zhang-xx-ML-GEC} developed by the authors of this paper, further provides a framework that is more convenient to analyze. The ML-GEC extends the ML-VAMP to cover an even broader scope of system models, by introducing some new entities to the message updating process. Inspired by these two works, this paper proposes a new algorithm that can efficiently detect the modulated symbols of massive MIMO relays when the number of antennas is median. The algorithm proposed is in essence a hybrid of the ML-VAMP and the ML-GEC, with certain particular optimization for the relay massive case. In an iterative manner it deliver message from each hop to the entire network and fulfil in effect a joint detection of all hops. As a hybrid, the algorithm enjoys many common superiorities from the previous two. One remarkable superiority is a good efficient-effective tradeof. The mean squared error (MSE) of the algorithm's output is Bayesian optimal while the computational complexity is under $O(n^3)$ per iteration, with $n$ denoting the antenna numbers. Another superiority is the theoretical predictability of the algorithm's asymptotic behavior. In large dimension, its MSE can be exactly predicted through the recursion of certain one-dimension equations, termed state evolution. It is also worthy of noting that, the algorithm, taking dual-hop for illustration, is readily extendable to the general case of multi-hop. Furthermore, when considering single-hop, it degenerates smoothly to the well-known results of EP \cite{Martin-TCOM14-EP} and VAMP \cite{Rangan-arxiv16-VAMP}.

Notations: $(\cdot)^H$ refers to conjugate transpose. $\propto$ means proportional to. $\text{Diag}(\bs{v})$ refers to the diagonal matrix with $\bs{v}$ on its diagonal positions, while $\text{diag}(\bs{A})$ is a vector of the diagonal elements of $\bs{A}$. $\odot$ denotes componentwise multiplication, while $\oslash$ for division. $\mathcal{N}_c(\bs{x} | \bs{\mu},\bs{\Sigma})$ denotes the complex Gaussian density with argument $\bs{x}$, mean $\bs{\mu}$, and covariance $\bs{\Sigma}$. We also abuse the notation $\mathcal{N}_c(\bs{x} | \bs{\mu},\bs{\sigma})$ to a Gaussian density whose covariance is a diagonal matrix with $\bs{\sigma}$ on the diagonal.
\begin{subequations}
\label{eq:def_Mean_Var}
\begin{align}
\mathbb{E} [ \bs{x} \,|\, \bs{m}, \bs{v}, \mathcal{F}(\cdot)]
    = &
    \frac{
        \int
            \bs{x}
            \mathcal{N} (\bs{x} | \bs{m}, \bs{v})
            \mathcal{F}(\bs{x})
        \, \mathrm{d} \bs{x}
    }{
        \int
            \mathcal{N} (\bs{x} | \bs{m}, \bs{v})
            \mathcal{F}(\bs{x})
        \, \mathrm{d} \bs{x}
    }
\\
\mathrm{Var} [ \bs{x} \,|\, \bs{m}, \bs{v}, \mathcal{F}(\cdot)]
    = &
    \frac{
        \int
            \bs{x} \bs{x}^H
            \mathcal{N} (\bs{x} | \bs{m}, \bs{v})
            \mathcal{F}(\bs{x})
        \, \mathrm{d} \bs{x}
    }{
        \int
            \mathcal{N} (\bs{x} | \bs{m}, \bs{v})
            \mathcal{F}(\bs{x})
        \, \mathrm{d} \bs{x}
    }
    -
\nonumber\\&
    \mathbb{E} [ \bs{x} \,|\, \bs{m}, \bs{v}, \mathcal{F}(\cdot)]
    \cdot
    \mathbb{E}^H [ \bs{x} \,|\, \bs{m}, \bs{v}, \mathcal{F}(\cdot)]
\end{align}

\end{subequations}

%%%%%%%%%%%%%%%%%%%%%%%%%%%%%%%%%%%%%%%%%%%%%%%%%%%%%%%%%%%%%%%%
\section{System Model}
%%%%%%%%%%%%%%%%%%%%%%%%%%%%%%%%%%%%%%%%%%%%%%%%%%%%%%%%%%%%%%%%
In this paper, we consider massive MIMO AF relays using a median number of antennas. We take dual-hop as an illustrating example and the system model is
\begin{align}
\mathrm{First \; hop: \;\;}
\bs{y}
&=
\mathcal{Q}  (\bs{Hx}+\bs{w}??)
\\
\mathrm{Second \; hop: \;\;}
\bs{z}
&=
\bs{Cy}+\bs{n}
\end{align}
where $\bs{x} \in \mathbb{C}^{L}$, $\bs{y} \in \mathbb{C}^{M}$, and $\bs{z} \in \mathbb{C}^{N}$ are the symbol vectors sent from the source node, repeated by the relay node, and received at the destination node, respectively. $\bs{H}\in \mathbb{C}^{M \times L}$ and $\bs{C}\in \mathbb{C}^{N \times M}$ are the channel matrices of the 1st and the 2nd hops respectively. $\bs{w} \in \mathbb{C}^{M}$ and $\bs{n} \in \mathbb{C}^{N}$ are the additive white Gaussian noise (AWGN), with $\bs{w}\sim \mathcal{N}_c(\bs{0},\sigma_1^2\bs{I})$ and $\bs{n}\sim \mathcal{N}_c(\bs{0},\sigma_2^2\bs{I})$ being known to the destination node.
The elements of the random vector $\bs{x}$ are independent and non-sparse (as opposed to compressed sensing), where $p(\bs{x})= \prod_{i=1}^M p_i(x_i)$. Given this non-sparsity, an additional condition is imposed on the dimensions, $L \leq M \leq N$, to make the detection non-trivial.
In AF relays, the vector $\bs{z}$ is observed (known), while the interim result $\bs{y}$ is hidden (unknown) from the destination node. We also assume the destination node to have perfect channel state information (CSI) about both hops. In practice, this CSI can be obtained by using channel estimators particularly designed for MIMO AF relays  \cite{Lioliou-JSAC12-RelayChanEst}.

In this paper, we aim at getting a Bayesian-optimal posterior mean estimate (PME) (optimal in the Bayesian sense) for each element $x_i$ of $\bs{x}$, given the observation $\bs{z}$, the channels, and the noise variances:
%本文关心的问题是Posterior mean estimate (PME), which was optimal in the sense of minimizing the output's MSE.
\begin{eqnarray}
\hat{x}_i
    & = &
    \mathbb{E}_{x_i | \bs{z}} [x_i]
    \label{eq:problem_states}
\end{eqnarray}
where the expectation is taken w.r.t. a marginal posterior density, $p ( x_i | \bs{z} )$, defined as
\begin{align}
p ( x_i | \bs{z} )
    & \propto
    \int_{\bs{x}_{\sim i}  }
    \!\!\!\!\!\!
    \mathrm{d}
    \bs{x}_{\sim i}
    \int_{\bs{t}}
    \mathrm{d} \bs{t}
    \int_{\bs{y}}
    p ( \bs{x} )
    %\cdot
    p ( \bs{t}|\bs{x} )
    %\cdot
    p ( \bs{y}|\bs{t} )
    %\cdot
    p ( \bs{z}|\bs{y} )
    \,
    \mathrm{d} \bs{y}
    \label{eq:posterior}
\end{align}
with
$
    p ( \bs{t}|\bs{x} )
    =
    \delta(\bs{t} - \bs{H} \bs{x})
$,
$
    p ( \bs{y}|\bs{t} )
    =
    \mathcal{N}_c (\bs{y}; \bs{t}, \sigma_1^2 \bs{I})
$, and
$
    p ( \bs{z}|\bs{y} )
    =
    \mathcal{N}_c (\bs{z}; \bs{C} \bs{y}, \sigma_2^2 \bs{I})
$.
The main difficulty in evaluating the multi-fold integral comes from two facts, first the high dimension of the vectors (i.e., antenna numbers), and second the large cardinality of the integral domain (i.e., the modulation order). The overall complexity is on the order of cardinality to dimension, which grows exponentially fast and makes the direct computation almost impossible.
An alternative solution to this problem is to use factor graph and message passing.

\section{The Proposed Algorithm}

\subsection{Algorithm}

\begin{figure}[!t]
\centering
  \includegraphics[angle=0,width=0.5\textwidth]{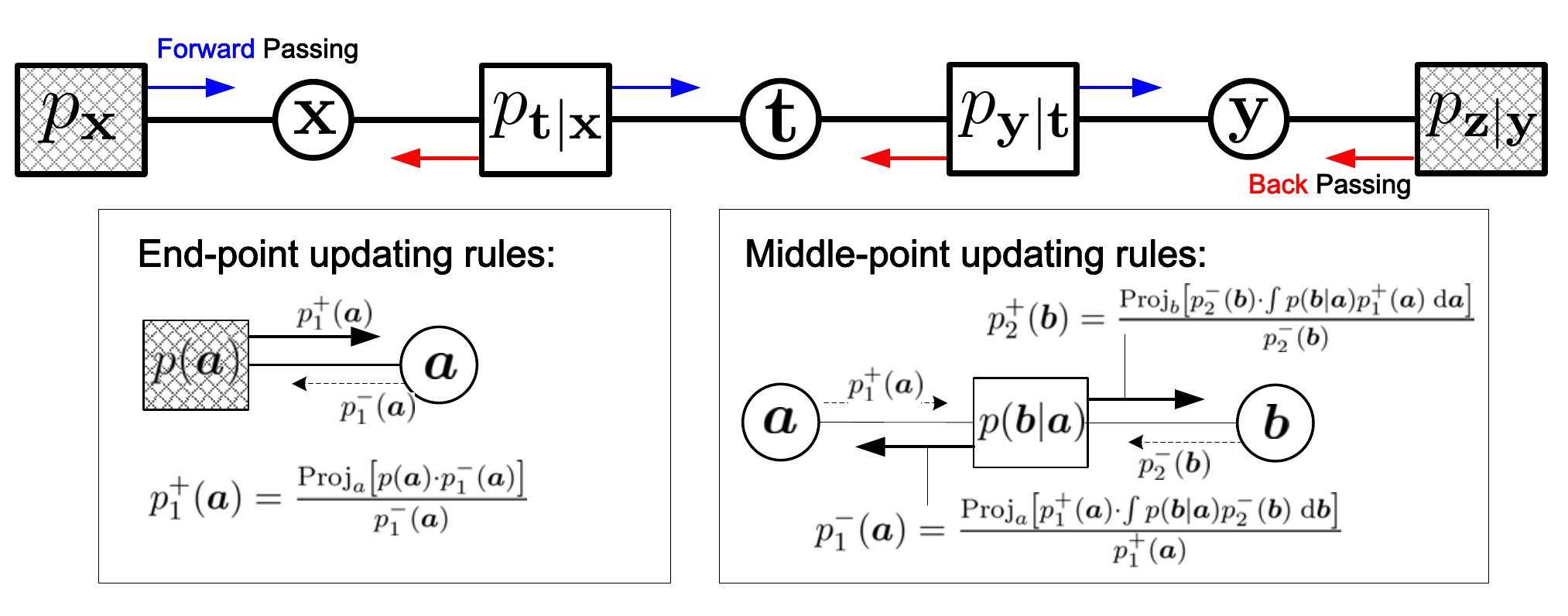}
  \caption{Factor graph in vector form and message updating rules}
  \label{fig:FG&MP}
\end{figure}

A factor graph associated with the density (\ref{eq:posterior}) is given in Fig. \ref{fig:FG&MP}.  There are two kinds of nodes in the figure,  the  variable nodes, and the factor nodes. Both nodes are in vector form, with a detailed description for the message updating also in Fig. \ref{fig:FG&MP}. Here, we note that the variable nodes  are transparent to the message passing. By contrast, a factor node will ``combine'' all messages received (from both directions) with its own (inherited from the factor function) ,  get a joint  ``information,''  subtract from it the incoming message in each direction by turns, and finally feeds back the ``extrinsic'' information to each direction. We call the delivery of a message from the left to the right in the factor graph a ``forward passing,'' and its opposite the ``back passing.'' Connecting a forward and a back passing, ones gets an iteration of the algorithm.  It is also worthy of noting that the messages delivered are indeed Gaussian densities, which means only the mean and the variance are needed. To approximate a general function $q(\cdot)$ by a Gaussian density, we adopt the same projection technique as \cite{Meng-SPL15-deriveAMP}, where
\begin{eqnarray}
\mathrm{Proj}_{\bs{x}} [q(\bs{x})]
    & = &
    \underset{ p(\bs{x}) \in \Phi }{\arg \min} \mathcal{D}_{\mathrm{KL}} (q(\bs{x}) ||  p(\bs{x}))
\end{eqnarray}
with $\Phi$ being the set of Gaussian densities in $\bs{x}$, and $\mathcal{D}_{\mathrm{KL}} (q || p)$ being the Kullback-Leibler divergence, a measure of how one probability distribution $q(\bs{x})$, diverges from a second, expected probability distribution $p(\bs{x})$. Based on this factor graph and its message passing rules, we proposed a new method, Algorithm \ref{alg:proposed}, to solve the PME problem in (\ref{eq:problem_states}). Some key steps of the derivations are given in Appendix \ref{app:derivation}. Here we note that the expectation in (\ref{eq:mod_D_1}) is taken w.r.t. a density proportional to $p(\bs{x}) \mathcal{N}_c (\bs{x} | \bs{m}_0^-, \bs{v}_0^-)$, and so is the variance in (\ref{eq:mod_D_2}).

\begin{algorithm}
\caption{The proposed algorithm} \label{alg:proposed}
\begin{algorithmic}[0] %每行显示行号
\State 1. Initialization: $\bs{v}_2^+=\bs{1}$, $\bs{v}_1^+=\bs{1}$, $\bs{v}_0^+=\bs{1}$, $\bs{m}_2^+=\bs{0}$, $\bs{m}_1^+=\bs{0}$, and $\bs{m}_0^+=\bs{0}$.
\State 2. Iteration (for $k=1\cdots, K$)
\State  \quad (1) Back passing
\begin{align}
\bs{Q}_y^-&=\left({\sigma_2^{-2}\bs{C}^H\bs{C}+ \text{Diag}(\bs{1} \oslash \bs{v}_2^+)}\right)^{-1}
\label{eq:Q_y^-}
\\
\hat{\bs{y}}^-&=\bs{Q}_y^- \left({\sigma_2^{-2}\bs{C}^H\bs{z}+\bs{m}_2^+ \oslash \bs{v}_2^+ }\right)
\label{eq:mod_A_1}
\\
\bs{v}_y^-&=\text{diag}(\bs{Q}_y^-).
\label{eq:mod_A_2}
\\
\bs{v}_2^-&=\bs{1}\oslash \left({\bs{1} \oslash \bs{v}_y^- - \bs{1}\oslash \bs{v}_2^+}\right)
\\
\bs{m}_2^-&=\bs{v}_2^- \odot \left({\hat{\bs{y}}^- \oslash \bs{v}_y^- - \bs{m}_2^+\oslash \bs{v}_2^+}\right)
\\
\bs{v}_t^-&=\bs{1} \oslash \left({\bs{1}\oslash \bs{v}_1^+ + \bs{1}\oslash (\sigma_1^2\bs{1}+\bs{v}_2^-)}\right)
\label{eq:mod_B_1}
\\
\hat{\bs{t}}^-&=\bs{v}_t^- \odot \left({ \bs{m}_1^+\oslash \bs{v}_1^+ +\bs{m}_2^- \oslash (\sigma_1^2\bs{1}+\bs{v}_2^-)}\right)
\label{eq:mod_B_2}
\\
\bs{v}_1^-&=\bs{1}\oslash \left({\bs{1} \oslash \bs{v}_t^- - \bs{1}\oslash \bs{v}_1^+}\right)
\\
\bs{m}_1^-&=\bs{v}_1^- \odot \left({\hat{\bs{t}}^- \oslash \bs{v}_t^- - \bs{m}_1^+\oslash \bs{v}_1^+}\right)
\\
\bs{Q}_x^-&=\left({\bs{H}^H\text{Diag}(\bs{1}\oslash \bs{v}_1^-)\bs{H}+\text{Diag}(\bs{1} \oslash \bs{v}_0^+)}\right)^{-1}
\label{eq:Q_x^-}
\\
\hat{\bs{x}}^-&=\bs{Q}_x^- \left({\bs{H}^H\text{Diag}(\bs{1}\oslash \bs{v}_1^-)\bs{m}_1^- + \bs{m}_0^+\oslash \bs{v}_0^+}\right)
\label{eq:mod_C_1}
\\
\bs{v}_x^-&=\text{diag}(\bs{Q}_x^-)
\label{eq:mod_C_2}
\\
\bs{v}_0^-&=\bs{1}\oslash \left({\bs{1} \oslash \bs{v}_x^- - \bs{1}\oslash \bs{v}_0^+}\right)
\\
\bs{m}_0^-&=\bs{v}_0^- \odot \left({\hat{\bs{x}}^- \oslash \bs{v}_x^- - \bs{m}_0^+\oslash \bs{v}_0^+}\right)
\end{align}
 \quad (2) Forward Passing
\begin{align}
\hat{\bs{x}}^+&=\mathbb{E}\left[{\bs{x}|\bs{m}_0^-,\bs{v}_0^-}\right]
\label{eq:mod_D_1}
\\
\bs{v}_x^+&=\text{Var}\left[{\bs{x}|\bs{m}_0^-,\bs{v}_0^-}\right]
\label{eq:mod_D_2}
\\
\bs{v}_0^+&=\bs{1}\oslash \left({\bs{1} \oslash \bs{v}_x^+ - \bs{1}\oslash \bs{v}_0^-}\right)
\\
\bs{m}_0^+&=\bs{v}_0^+ \odot \left({\hat{\bs{x}}^+ \oslash \bs{v}_x^+ - \bs{m}_0^-\oslash \bs{v}_0^-}\right)
\\
\bs{Q}_x^+&=\left({\bs{H}^H\text{Diag}(\bs{1}\oslash \bs{v}_1^-)\bs{H}+\text{Diag}(\bs{1}\oslash \bs{v}_0^+)}\right)^{-1}
\label{eq:Q_x^+}
\\
\bs{x}^+&=\bs{Q}_x^+ \left({\bs{H}^H\text{Diag}(\bs{1}\oslash \bs{v}_1^-)\bs{m}_1^- + \bs{m}_0^+ \oslash \bs{v}_0^+}\right)
\\
\hat{\bs{t}}^+&=\bs{H}\bs{x}^+
\label{eq:mod_C_3}
\\
\bs{v}_t^+&=\text{diag}(\boldsymbol{H}\boldsymbol{Q}_x\boldsymbol{H}^H)
\label{eq:mod_C_4}
\\
\bs{v}_1^+&=\bs{1}\oslash \left({\bs{1} \oslash \bs{v}_t^+ - \bs{1}\oslash \bs{v}_1^-}\right)
\\
\bs{m}_1^+&=\bs{v}_1^+ \odot \left({\hat{\bs{t}}^+ \oslash \bs{v}_t^+ - \bs{m}_1^-\oslash \bs{v}_1^-}\right)
\\
\bs{v}_y^+&=\bs{1}\oslash \left({(\bs{1} \oslash \bs{v}_2^-) +\bs{1}\oslash (\bs{v}_1^++\sigma_1^2\bs{1})}\right)
\label{eq:mod_B_3}
\\
\hat{\bs{y}}^+&=\bs{v}_y^+\odot \left({\bs{m}_1^+ \oslash (\bs{v}_1^++\sigma_1^2\bs{1})+\bs{m}_2^-\oslash \bs{v}_2^-}\right)
\label{eq:mod_B_4}
\\
\bs{v}_2^+&=\bs{1}\oslash \left({\bs{1} \oslash \bs{v}_y^+ - \bs{1}\oslash \bs{v}_2^-}\right)
\\
\bs{m}_2^+&=\bs{v}_2^+ \odot \left({\hat{\bs{y}}^+ \oslash \bs{v}_y^+ - \bs{m}_2^-\oslash \bs{v}_2^-}\right)
\end{align}
\State 3. Output: $\hat{\bs{x}}^+$.

\end{algorithmic}
\end{algorithm}

\subsection{Block Diagram}
\begin{figure}[!t]
\centering
  \includegraphics[width=0.5\textwidth]{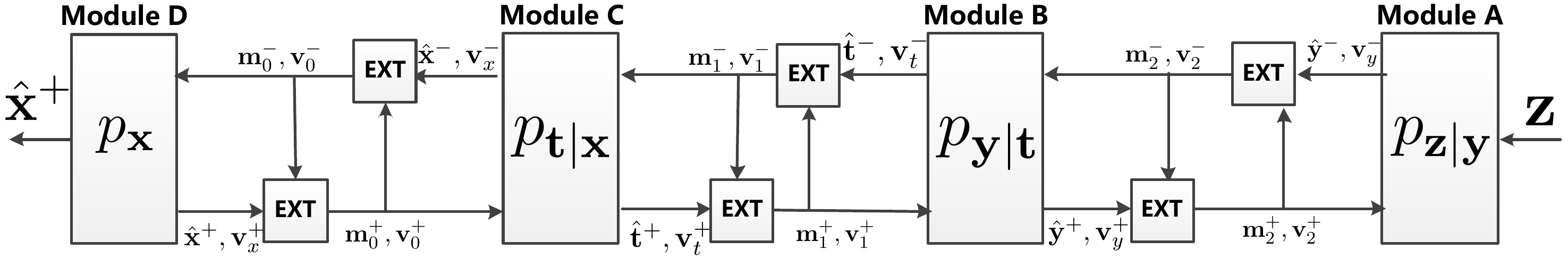}
  \caption{Block diagram of the proposed algorithm}
  \label{fig:block_diagram}
\end{figure}
To better illustrate the working mechanism of our algorithm, also to shed light on the software and hardware implementation, we provide in Fig. \ref{fig:block_diagram} a block diagram of the algorithm. In the diagram, Module A corresponds to (\ref{eq:mod_A_1})- (\ref{eq:mod_A_2}) of Algorithm \ref{alg:proposed}, with Module B to (\ref{eq:mod_B_1})-(\ref{eq:mod_B_2}) and  (\ref{eq:mod_B_3})-(\ref{eq:mod_B_4}), Module C to  (\ref{eq:mod_C_1})-(\ref{eq:mod_C_2}) and  (\ref{eq:mod_C_3})-(\ref{eq:mod_C_4}), and Module D to (\ref{eq:mod_D_1})-(\ref{eq:mod_D_2}). Some discussions on the diagram are also given below.
\begin{itemize}
\item
    The algorithm proposed, indeed, provides an iterative manner of jointly estimating the 1st and the 2nd hops, with $\hat{\bs{x}}$ and $\hat{\bs{y}}$ as the estimated results. The rationale behind a joint estimation to outperform algorithms which handle each hop separately is that more information can be utilized by each part of the network, and a global optima is more likely to be attained. The superiority of our joint processing will be discussed with more details in Section \ref{sec:simulation}.

\item
    A key step in our iterative algorithm is to compute the \emph{extrinsic}  message, represented by $ \boxed{\mathrm{EXT}} $. This operation bears a similarity to the turbo-like processing, which allows only the extrinsic information to traverse the network, suppressing error propagation to the minimum. The computation is formally defined as
$
(\bs{m}_1, \bs{v}_1) \,
\boxed{\mathrm{EXT}} \,
(\bs{m}_2, \bs{v}_2)
     =
    \left[
        \bs{v}_3 \odot ( \bs{m}_1 \oslash \bs{v}_1 - \bs{m}_2 \oslash \bs{v}_2 )
        ,
        \bs{v}_3
    \right]
$, with
$\bs{v}_3 = \bs{1} \oslash (\bs{1} \oslash \bs{v}_1 - \bs{1} \oslash \bs{v}_2 )$.
%这个subraction的操作是整个算法能正常工作的关键步骤，其原理与turbo code 译码过程中计算extrinsic information相似，目的都是为了避免由正反馈所引起的错误传播。

\item
  Most computational burden of the proposed algorithm comes from Module C in the diagram. To be specific, the matrix inversions in (\ref{eq:Q_y^-}), (\ref{eq:Q_x^-}), and (\ref{eq:Q_x^+}) of Algorithm \ref{alg:proposed} all have a complexity on the order of $O(n^3)$, with $n$ denoting the dimension of the matrix involved. This complexity, originating from EP \cite{Minka-PhD01-EP}, seems to be inherited by ML-VAMP \cite{Fletcher-arxiv17-ML-VAMP} and ML-GEC \cite{Zhang-xx-ML-GEC}. Fortunately, a new approach to avoid the matrix inversion has been developed in \cite{Cakmak-arxiv18-EPDiagRest} very recently. That approach is also applicable to the algorithm here, but we save it for further studies.

\item
    Given the block diagram, we now explain why the proposed algorithm is a hybrid of the ML-VAMP \cite{Fletcher-arxiv17-ML-VAMP} and the ML-GEC \cite{Zhang-xx-ML-GEC} with particular optimization for relay massive. There are at least three facets. First, the proposed algorithm removes the (deep learning) activation functions from the ML-VAMP since it is not needed here. Second, after the removal above, it follows the same procedure as the ML-GEC to update the message of most factor nodes except the last two. Third, it merges the ML-GEC's last two factors nodes to be a single one, i.e., Module A in Fig. \ref{fig:block_diagram}, shortening the entire message passing route.

\item
    Relay or multi-hop communication is a general framework that embraces the single-hop. It is worthy of noting our algorithm can be generalized directly to cover multi-hop, although Fig. \ref{fig:block_diagram} only gives a dual-hop example. The generation to $m$-hop is done by first creating $m-1$ copies of the Module B and C pair, then changing the channel matrix and noise variance for each pair accordingly, and finally concatenating them in ascending order of the hop indices. We also note that our algorithm degenerates easily to the well-known results for single-hop. To see this, one only needs to take out Module B and C. After that, another algorithm (block diagram also)  is immediately seen, which is exactly the same as EP \cite{Martin-TCOM14-EP} and VAMP \cite{Rangan-arxiv16-VAMP}.
\end{itemize}

\subsection{State Evolution}
Inheriting from the AMP and the EP classes,  the proposed algorithm admits an exact description of the high-dimensional behavior. Specifically, the variance of the output can be computed through a one-dimensional recursion, i.e., the SE.

To this end, a scalar system model and some new parameters are introduced. The scalar model is:
$%\begin{align}
r=x+w
%\label{equ1}
$, %\end{align}
with $w\sim \mathcal{N}_c\left({0,v_0^-}\right)$.  The PME output then becomes
$%\begin{align}
\mathbb{E}\left[{x|r}\right]=\int_{x} xp(x)p(r|x)\, \mathrm{d}x
$, %\end{align}
with the MSE given by
$%\begin{align}
\text{MSE}(v_0^-)=\mathbb{E}\left[{|x-\mathbb{E}\left[{x|r}\right]|^2}\right]
$, %\end{align}
where the expectation is taken w.r.t. $p(r,x)=p(x)p(r|x)$. We denote $\lambda_i$ as the $i$-th eigenvalue of $\bs{H}^H\bs{H}$, and  $\eta_i$ as that of $\bs{C}^H\bs{C}$. We also define $ \langle f(\lambda_i) \rangle_N \triangleq \frac{1}{N} \sum_{i=1}^N f(\lambda_i)$.
Then, we derive the SE for the proposed algorithm. The results are given in Algorithm \ref{alg:SE}. Details of the derivation are omitted here. Similar techniques as in \cite{He-ISIT17-GEC-SR} can be applied to the current case, while one has to go through the derivations from the beginning.

%
%\begin{align}
%\gamma_0^+=\frac{1}{v_0^+}, \ \gamma_1^+=\frac{1}{v_1^+},\ \gamma_2^+=\frac{1}{v_2^+},\ \gamma_y^+=\frac{1}{v_y^+},
%\gamma_0^-=\frac{1}{v_0^-}, \ \gamma_1^-=\frac{1}{v_1^-},\ \gamma_2^-=\frac{1}{v_2^-},\ \gamma_t^-=\frac{1}{v_t^-}
%\end{align}
%
%In addition, $\lambda_i$ in algorithm \ref{A2} is the $i$-th eigenvalue of $\bs{H}^H\bs{H}$, while $\eta_i$ is the $i$-th of $\bs{C}^H\bs{C}$. The expectation with respect to $\lambda$ is defined by
%\begin{align}
%\mathbb{E}[f(\lambda)]=\frac{1}{N}\sum\limits_{i=1}^Nf(\lambda_i)
%\end{align}
%In order to get $q_x^-$, we obtain the SVD of matrix $\bs{H}$.
%\begin{align}
%\bs{H}=\bs{U}\bs{\Sigma}\bs{V}
%\end{align}
%Then, $\bs{Q}_x^-$ can be rewritten as
%\begin{align}
%\bs{Q}_x^-=\bs{V}\left({\bs{\Sigma}\text{Diag}(\bs{1}\oslash \bs{v}_1^-)\bs{\Sigma}^H+\text{Diag}(\bs{1} \oslash \bs{v}_0^+)}\right)^{-1}\bs{V}^H
%\end{align}
%The diagonal elements of $\bs{Q}_{x}^-$ is given by
%\begin{align}
%q_x=\mathbb{E}\left[{\frac{1}{\lambda_i \gamma_1^- +\gamma_0^+}}\right]=\frac{1}{L}\sum\limits_{i=1}^L \left({\frac{1}{ \frac{\lambda_i}{v_1^-} +\frac{1}{v_0^+}} }\right)
%\end{align}
%Similarly, $q_y^-$ and $q_t^+$ also can be obtained as follows
%\begin{align}
%q_y^-=\mathbb{E}\left[{\frac{1}{\sigma_2^{-2}\eta_i+\gamma_2^+}}\right], \quad q_t^+=\mathbb{E}\left[\frac{\lambda_i}{\lambda_i\gamma_1^- +\gamma_0^+}\right]
%\end{align}

\begin{algorithm}
\caption{State Evolution}
\label{alg:SE}
\begin{algorithmic}[0] %每行显示行号
\State
1. Initial: $k$=0, $\gamma_2^+=1$, $\gamma_1^+=1$, $\gamma_0^+=1$\\
2. Iteration (for $k=1:K$)
\State \quad Back Passing
\begin{align}
q_y^-
&=\langle {1}/{(\sigma_2^{-2}\eta_i+\gamma_2^+)} \rangle_M
\\
\gamma_2^-
&={1}/{q_y^-}-\gamma_2^+
\\
\gamma_t^-
&=\gamma_1^+ + {\gamma_2^-}/{(1+\gamma_2^-\sigma_1^{2})}
\\
\gamma_1^-
&=\gamma_t^- -\gamma_1^+
\\
q_x^-
&=\langle {1}/{(\lambda_i \gamma_1^- +\gamma_0^+)} \rangle_M
\\
\gamma_0^-
&=\frac{1}{q_x^-}-\gamma_0^+
\end{align}
\quad Forward Passing
\begin{align}
\gamma_0^+&={1}/{\text{MSE}(\gamma_0^-)} -\gamma_0^-\\
q_t^+&= \langle {\lambda_i}/{(\lambda_i\gamma_1^- +\gamma_0^+)} \rangle_L
\\
\gamma_1^+&={1}/{q_t^+}-\gamma_1^-\\
\gamma_y^+&=\gamma_2^- +{\gamma_1^+}/{(1+\gamma_1^+\sigma_1^2)}\\
\gamma_2^+&=\gamma_y^+ -\gamma_2^-
\end{align}
3. Output: $\text{MSE}(\gamma_0^-)$
\end{algorithmic}
\end{algorithm}

\section{Numerical Results} \label{sec:simulation}
%为了验证本文所提算法的有效性，我们将它与另外3种算法同时进行Monte Carlo 仿真，并比较他们之间性能的差异。这三种算法是：
In this section, we carry out numerical simulations and compare the results with other competing methods to verify the effectiveness of Algorithm \ref{alg:proposed}. Three algorithms, upon which linear/nonlinear and separate-/joint-processing are applied, are chosen to differentiate from our iterative (nonlinear) joint estimation%
\footnote{
As mentioned earlier, our algorithm is a hybrid of the ML-VAMP and the ML-GEC, with particular optimization to a shorten the message passing route. The computational complexity is reduced after the removal and mergence of factor nodes; however, it remains on the same same order (due to the same bottleneck of matrix inversion). The BER/MSE performance of the three is also similar. For this reason, we omit the comparison among the three, but focusing more on their difference to techniques in other class.
}%
. These three are:
\begin{itemize}
    \item
        {\bf LMMSE+LS} (linear, separate): Perform LS estimation on the 2nd hop,  and use the result as the 1st hop's observation, to later perform an LMMSE estimation on the 1st hop (Performing LMMSE also on the 2nd hop is difficult due to the lack of $\bs{y}$'s prior information).
    \item
        {\bf Single-LMMSE} (linear, joint):  Instead of separating the two hops, use a compound model of $\bs{z} = \bs{C} \bs{H} \bs{x} + \bs{C} \bs{w} + \bs{n}$, and simply treat the colored noise $\bs{C} \bs{w} + \bs{n}$ as a white one with equivalent variance, then perform a single LMMSE estimation over the entire link.
    \item
        {\bf EP+LS} (nonlinear, separate): Similar to the LMMSE+LS above, with LS replaced with EP \cite{Martin-TCOM14-EP}, which is  nonlinear.
\end{itemize}

In the numerical simulations, we fix the antenna numbers of the source, relay, and destination at $L=128$, $M=256$, and $N=512$, respectively. Uncorrelated flat Rayleigh fading channels and QPSK without channel coding are assumed. The SNR of the 1st and 2nd hops are denoted as $\mathrm{SNR}_1$ and $\mathrm{SNR}_2$, respectively.

\begin{figure}[!t]
\centering
\includegraphics[width=0.45\textwidth]{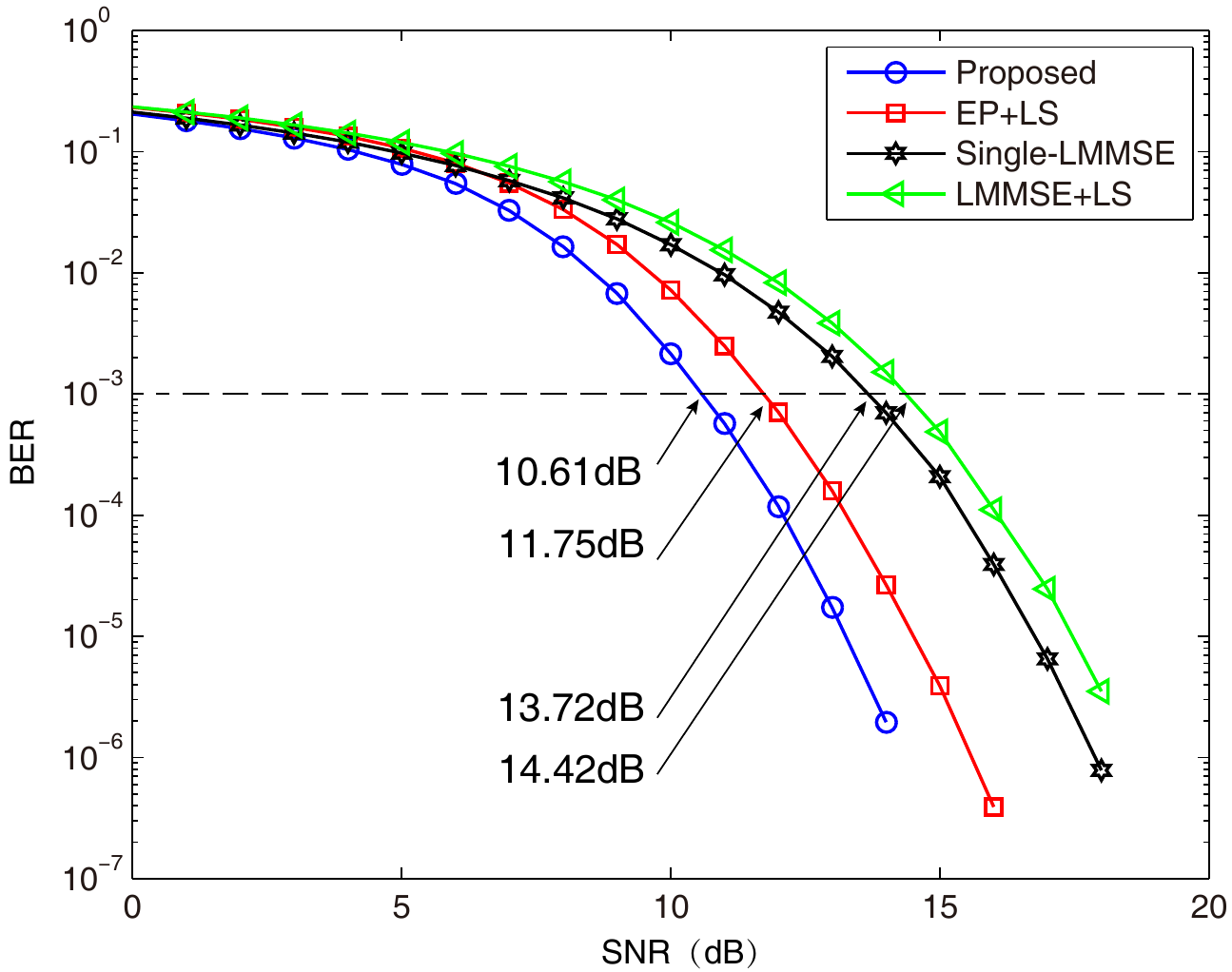}
\caption{Comparison of BER performance ($\mathrm{SNR}_2 = \mathrm{SNR}_1 - 3 \mathrm{dB}$).}
\label{fig:Fix_3dB}
\end{figure}

Fig. \ref{fig:Fix_3dB} compares the bit error rate (BER) of all the four algorithms. Clearly, the proposed algorithm outperforms the other three in BER over the entire SNR range. LMMSE+LS is the worst, with Single-LMMSE only slightly better. These two algorithms both suffer from the non-optimality of the LMMSE, and thus, it is not surprising that, replacing LMMSE with EP (proved to be a better technique \cite{Martin-TCOM14-EP}) yields a much better algorithm. EP+LS, however, is still inferior to the proposed algorithm. This inferiority could be attributed to the noise augmentation of the LS method in the 2nd hop.

\begin{figure}[!t]
\centering
\includegraphics[width=0.45\textwidth]{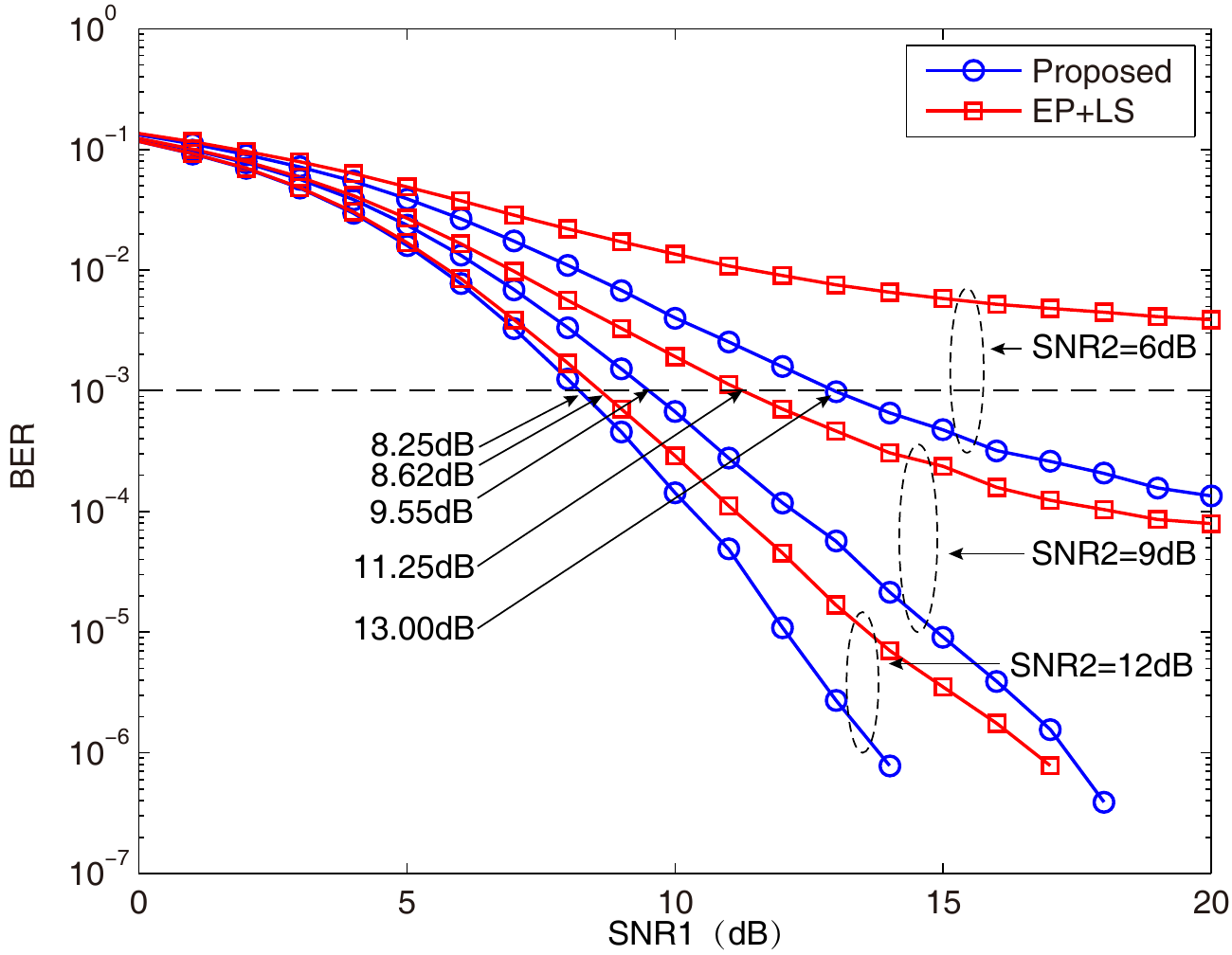}
\caption{BER of the EP+LS and the proposed algorithm}
\label{fig:Fix_SNR2}
\end{figure}

To gain more insights into the proposed algorithm's behavior, we fix the $\mathrm{SNR}_2$ at three levels, i.e., 6dB, 9dB, and 12dB, and vary the $\mathrm{SNR}_1$ from 0 to 20 dB. The BER verse $\mathrm{SNR}_1$ results are presented in Fig. \ref{fig:Fix_SNR2}. We find that, firstly, the proposed algorithm outperforms EP+LS in all cases and over the entire SNR range. Secondly, the difference between the two tends to diminish as the $\mathrm{SNR}_2$ soars up. This phenomenon also meets our expectation that as the $\mathrm{SNR}_2$ increases, the LS estimation is becoming more and the more accurate, leaving little room for improvement. In other words, the proposed algorithm differs from the EP+LS in the 2nd hop processing, where the proposed algorithm implicitly utilizes some statistical information about the 2nd hop's input $\bs{y}$ in the detection process, while the EP+LS has nothing in prior to depend on (except the channel matrix).

\begin{figure}[!t]
\centering
  \includegraphics[width=0.45\textwidth]{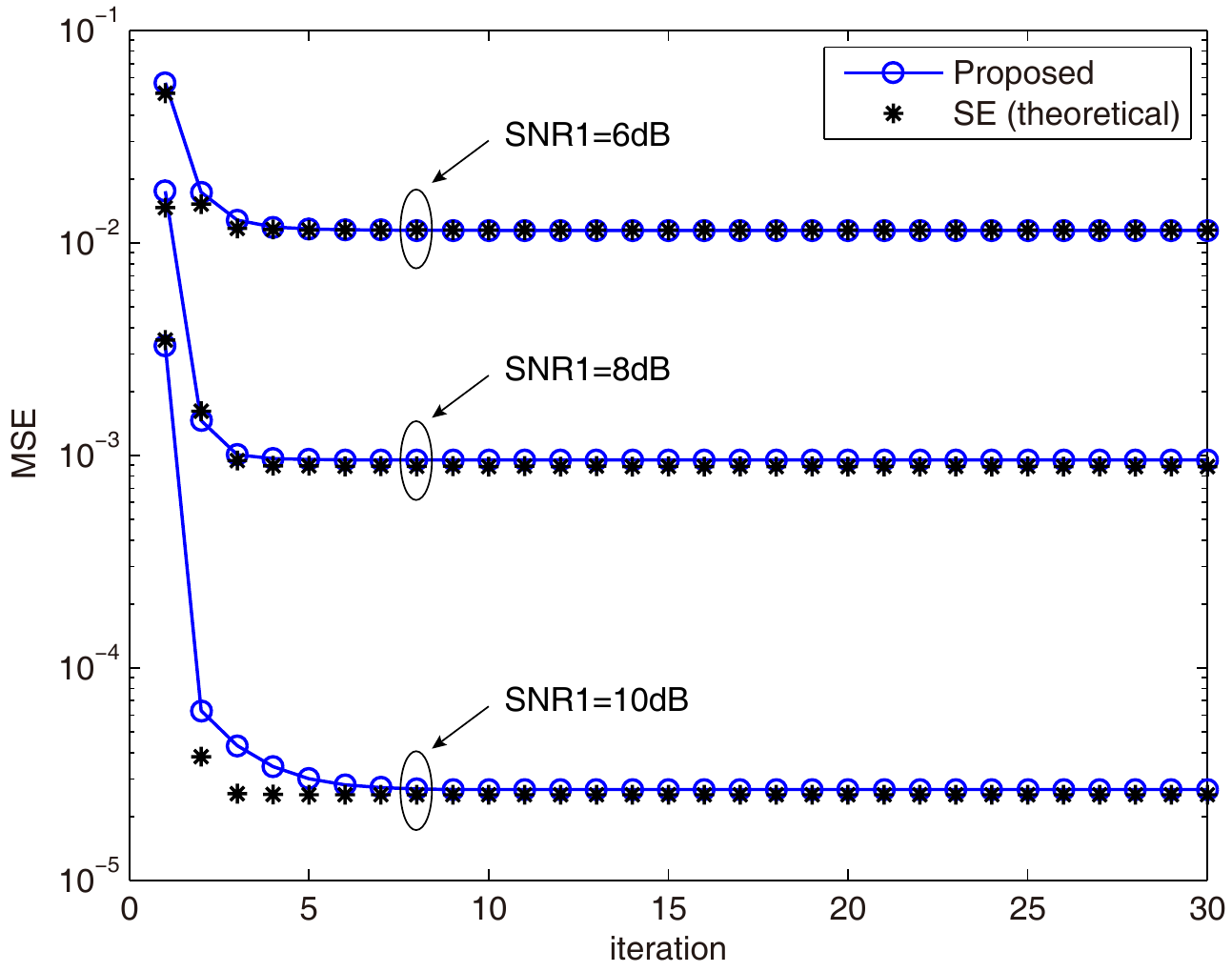}
  \caption{MSE of the proposed algorithm ($\mathrm{SNR}_2=20$dB)}
  \label{fig:Iter}
\end{figure}

In Fig. \ref{fig:Iter}, we provide a closer look at the proposed algorithm's output MSE per iteration. We find that the algorithm has a quick convergence. Only 5 iterations or less are needed in all cases simulated. Moreover, the algorithm's real MSE per iteration matches perfectly with the one theoretically obtained through state evolution.

\section{Conclusions}

This paper considered the problem of symbol detection in the context of massive MIMO AF relays using a median number of antennas. A new algorithm has been proposed to optimize the tradeoff between effectiveness and efficiency. It can attain Bayesian-optimal MSE at only the cost of $O(n^3)$  complexity. The algorithm is in essence a hybrid of ML-VAMP and ML-GEC, with particular optimization for the massive MIMO relay case. As a hybrid, it inherits from the two the formulism of state evolution, i.e., the asymptotic MSE behavior of the algorithm can be precisely predicted through a scalar equivalent model. Furthermore, it can smoothly degenerate to the well-known result of EP and VAMP when single-hop communication is considered.

\begin{appendices}
\section{General Derivation} \label{app:derivation}
%We start by computing the joint message of $\boldsymbol{y}$ (Module A in Fig. \ref{fig:block_diagram})
%\begin{eqnarray}
%%&&
%\mathcal{N}_c\left({\bs{y}|\hat{\bs{y}}^-,\text{Diag}(\bs{v}_y^-)}\right)
%%\nonumber\\
%&=%&
%\text{Proj}_{\bs{y}}\left[{\frac{p(\bs{z}|\bs{y})\mathcal{N}_c(\bs{y}|\bs{m}_2^+,\text{Diag}(\bs{v}_2^+))}{\int_{\bs{y}}p(\bs{z}|\bs{y})\mathcal{N}_c(\bs{y}|\bs{m}_2^+,\text{Diag}(\bs{v}_2^+))\, \mathrm{d}\bs{y}}}\right]
%\end{eqnarray}
%where
%\begin{align}
%\hat{\bs{y}}^-&=({ \sigma_2^{-2}\bs{C}^H\bs{C}+\text{Diag}(\bs{1}\oslash \bs{v}_2^+)})^{-1}
%({\sigma_2^{-2}\bs{C}^H\bs{z}+\bs{m}_2^+\oslash \bs{v}_2^+})\\
%\bs{v}_y^-&=\text{diag}\left[{ ({ \sigma_2^{-2}\bs{C}^H\bs{C}+\text{Diag}(\bs{1} \oslash \bs{v}_2^+)})^{-1} }\right]
%\end{align}
%where `$\oslash$' represents componentwise multiplication.
%
%Similar to $\bs{x}$, the joint message of $\bs{x}$ can be computed as follows (Module D in Fig. \ref{fig:block_diagram})
%\begin{align}
%\mathcal{N}_c (\bs{x}|\hat{\bs{x}}^+,\text{Diag}(\bs{v}_x^+))=\text{Proj}_{\bs{x}}\left[{\frac{p(\bs{x})\mathcal{N}_c(\bs{x}|\bs{m}_0^-,\text{Diag}(\bs{v}_0^-))}{\int_{\bs{x}}p(\bs{x})\mathcal{N}_c(\bs{x}|\bs{m}_0^-,\text{Diag}(\bs{v}_0^-))\, \mathrm{d}\bs{x}}}\right]
%\end{align}
%where
%\begin{align}
%\hat{\bs{x}}^+&=\mathbb{E}[\bs{x}|\bs{m}_0^-,\bs{v}_0^-]\\
%\bs{v}_x^+ &=\text{Var}[\bs{x}|\bs{m}_0^-,\bs{v}_0^-]
%\end{align}

\subsection{Back passing}
The belief distribution at the factor node $p({\bs{z} | \bs{y}}) = \mathcal{N}_c\left(\bs{C} \bs{y} | \bs{z}, \sigma_2^2 \bs{1}\right)$, i.e., Module A in Fig. \ref{fig:block_diagram}, is
\begin{align}
\mathcal{N}_c\left({\bs{y}|\hat{\bs{y}}^-, \bs{v}_y^-}\right)
=
\text{Proj}_{\bs{y}}\left[{p(\bs{z}|\bs{y})\mathcal{N}(\bs{y}|\bs{m}_2^+, \bs{v}_2^+)}\right]
\end{align}
which can be computed via the use of the Gaussian reproduction property \cite{Rasmussen-Book04-GaussianReprod}%
\footnote{
Gaussian reproduction property:
$\mathcal{N}_c\left({\bs{x}|\bs{a},\bs{A}}\right)\mathcal{N}_c\left({\bs{x}|\bs{b},\bs{B}}\right)
= \alpha \cdot \mathcal{N}_c\left({\bs{x}|\bs{c},\bs{C}}\right)
$, where $\alpha=\mathcal{N}_c\left({\bs{0}|\bs{a}-\bs{b},\bs{A}+\bs{B}}\right)$, $\bs{c}=(\bs{A}^{-1}+\bs{B}^{-1})^{-1}(\bs{A}^{-1}\bs{a}+\bs{B}^{-1}\bs{b})$, and $\bs{C}=(\bs{A}^{-1}+\bs{B}^{-1})^{-1}$.
}%
. The result is
\begin{align}
\hat{\bs{y}}^-&=\bs{Q}_y^- \left({\sigma_2^{-2}\bs{C}^H\bs{z}+\bs{m}_2^+ \oslash \bs{v}_2^+ }\right)\\
\bs{v}_y^-&=\text{diag}(\bs{Q}_y^-).
\end{align}
with
$
\bs{Q}_y^-
=
\left({\sigma_2^{-2}\bs{C}^H\bs{C}+ \text{Diag}(\bs{1} \oslash \bs{v}_2^+)}\right)^{-1}
$.
After that, the back-passing extrinsic message can be computed as
\begin{align}
\mathcal{N}_c\left({\bs{y}|\bs{m}_2^-, \bs{v}_2^-}\right)\propto
\frac{\mathcal{N}_c\left({\bs{y}|\hat{\bs{y}}^-, \bs{v}_y^- }\right)}{\mathcal{N}_c\left({\bs{y}|\bs{m}_2^+, \bs{v}_2^+}\right)}
\end{align}
with the result being
\begin{align}
\bs{v}_2^-&=\bs{1}\oslash \left({\bs{1} \oslash \bs{v}_y^- - \bs{1}\oslash \bs{v}_2^+}\right)\\
\bs{m}_2^-&=\bs{v}_2^- \odot \left({\hat{\bs{y}}^- \oslash \bs{v}_y^- - \bs{m}_2^+\oslash \bs{v}_2^+}\right)
\end{align}

We move on to the factor node $p(\bs{y} | \bs{t})$, i.e., Module B,
\begin{align}
\mathcal{N}_c\left({\bs{t}|\hat{\bs{t}}^-, \bs{v}_t^- }\right)
%\nonumber\\
%=&
%\text{Proj}_{\bs{t}}
%\left[
%    \mathcal{N}_c(\bs{t}|\bs{m}_1^+, \bs{v}_1^+)
%    \int_{\bs{y}}
%        p(\bs{y}|\bs{t})
%        \mathcal{N}_c(\bs{y}|\bs{m}_2^-, \bs{v}_2^-)
%    \,\mathrm{d}\bs{y}
%\right]
%\\
=&
\text{Proj}_{\bs{t}}
\left[
    \mathcal{N}_c(\bs{t}|\bs{m}_1^+, \bs{v}_1^+)
    \mathcal{L}(\bs{t})
\right]
\end{align}
where
$
\mathcal{L}(\bs{t}) \triangleq
\int_{\bs{y}} p(\bs{y}|\bs{t}) \mathcal{N}_c(\bs{y}|\bs{m}_2^-, \bs{v}_2^-)    \,\mathrm{d}\bs{y}
$,
and according to our notation systems (\ref{eq:def_Mean_Var}), the mean and variance can be expressed as
\begin{align}
\hat{\bs{t}}^-
&=
    \mathbb{E}
    \left[
        \bs{t} \,|\, \bs{m}_1^+, \bs{v}_1^+, \mathcal{L}(\cdot)
    \right]
\\
\bs{v}_t^-
&=
    \mathrm{Var}
    \left[
        \bs{t} \,|\, \bs{m}_1^+, \bs{v}_1^+, \mathcal{L}(\cdot)
    \right]
\end{align}
Then the extrinsic message can be obtained
\begin{align}
\mathcal{N}_c\left({\bs{t}|\bs{m}_1^-, \bs{v}_1^-}\right)
&\propto \frac{\mathcal{N}_c\left({\bs{t}|\hat{\bs{t}}^-, \bs{v}_t^-}\right)}{\mathcal{N}_c\left({\bs{t}|\bs{m}_1^+, \bs{v}_1^+}\right)}
\\
\bs{v}_1^-
&=\bs{1}\oslash \left({\bs{1} \oslash \bs{v}_t^- - \bs{1}\oslash \bs{v}_1^+}\right)
\\
\bs{m}_1^-
&=\bs{v}_1^- \odot \left({\hat{\bs{t}}^- \oslash \bs{v}_t^- - \bs{m}_1^+\oslash \bs{v}_1^+}\right)
\end{align}

We continue with the $p(\bs{t} | \bs{x}) = \delta(\bs{t}-\bs{Hx})$  factor node, i.e. Module C,
\begin{align}
& \mathcal{N}_c\left({\bs{x}|\hat{\bs{x}}^-, \bs{v}_x^-}\right)
\nonumber\\
= &
\text{Proj}_{\bs{x}}
\left[
\int_{\bs{t}}
p(\bs{t}|\bs{x})\mathcal{N}_c\left({\bs{x}|\bs{m}_0^+, \bs{v}_0^+}\right)
\mathcal{N}_c\left({\bs{t}|\bs{m}_1^-, \bs{v}_1^-}\right)
\, \mathrm{d} \bs{t}
\right]
\end{align}
Similarly, we obtain
\begin{align}
\bs{Q}_x^-&=\left({\bs{H}^H\text{Diag}(\bs{1}\oslash \bs{v}_1^-)\bs{H}+\text{Diag}(\bs{1} \oslash \bs{v}_0^+)}\right)^{-1}\\
\hat{\bs{x}}^-&=\bs{Q}_x^- \left({\bs{H}^H\text{Diag}(\bs{1}\oslash \bs{v}_1^-)\bs{r}_1^- + \bs{m}_0^+\oslash \bs{v}_0^+}\right)\\
\bs{v}_x^-&=\text{diag}(\bs{Q}_x^-)
\end{align}
Then, the extrinsic message becomes
\begin{align}
\mathcal{N}_c\left({\bs{x}|\bs{r}_0^-, \bs{v}_0^-}\right)
&\propto \frac{\mathcal{N}_c\left({\bs{x}|\hat{\bs{x}}^-, \bs{v}_x^-}\right)}{\mathcal{N}_c\left({\bs{x}|\bs{m}_0^+, \bs{v}_0^+}\right)}
\\
\bs{v}_0^-
&=\bs{1}\oslash \left({\bs{1} \oslash \bs{v}_x^- - \bs{1}\oslash \bs{v}_0^+}\right)
\\
\bs{r}_0^-
&=\bs{v}_0^- \odot \left({\hat{\bs{x}}^- \oslash \bs{v}_x^- - \bs{r}_0^+\oslash \bs{v}_0^+}\right)
\end{align}

\subsection{Forward passing}
We compute the joint message of the factor node $p(\bs{x})$, i.e., Module D in Fig. \ref{fig:block_diagram},
\begin{align}
\mathcal{N}_c\left({\bs{x}|\hat{\bs{x}}^+, \bs{v}_x^+}\right)
&=\text{Proj}_{\bs{x}}\left[{p(\bs{x})\mathcal{N}_c\left({\bs{x}|\bs{m}_0^-,\bs{v}_0^-}\right)}\right]
\\
\hat{\bs{x}}^+
&=\mathbb{E}\left[{\bs{x}|\bs{m}_0^-,\bs{v}_0^-}\right]
\\
\bs{v}_x^+
&=\text{diag}(\text{Var}\left[{\bs{x}|\bs{m}_0^-,\bs{v}_0^-}\right])
\end{align}
Then, the extrinsic message can be computed as
\begin{align}
\mathcal{N}_c(\bs{x}|\bs{m}_0^+, \bs{v}_0^+)
&\propto
\frac{\mathcal{N}_c\left({\bs{x}|\hat{\bs{x}}^+, \bs{v}_x^+}\right)}{\mathcal{N}_c(\bs{x}|\bs{m}_0^-, \bs{v}_0^-)}
\\
\bs{v}_0^+
&=\bs{1}\oslash \left({\bs{1} \oslash \bs{v}_x^+ - \bs{1}\oslash \bs{v}_0^-}\right)
\\
\bs{m}_0^+
&=\bs{v}_0^+ \odot \left({\hat{\bs{x}}^+ \oslash \bs{v}_x^+ - \bs{m}_0^-\oslash \bs{v}_0^-}\right)
\end{align}
We move on to the factor node $p(\bs{t} |\bs{x}) = \delta(\bs{t}-\bs{Hx})$, i.e., Module C,
\begin{align}
\mathcal{N}_c\left({\bs{t}|\hat{\bs{t}}^+, \bs{v}_t^+}\right)
=&
\text{Proj}_{\bs{t}}
[
\int_{\bs{x}}
p(\bs{t}|\bs{x})\mathcal{N}_c\left({\bs{x}|\bs{m}_0^+,\bs{v}_0^+}\right)
\nonumber\\
& \cdot
\mathcal{N}_c\left({\bs{t}|\bs{m}_1^-, \bs{v}_1^-}\right)
\mathrm{d} \bs{x}
]
\end{align}
The integral above can be computed as
\begin{align}
& \int_{\bs{x}}
        p(\bs{t}|\bs{x})
        \mathcal{N}_c\left({\bs{x}|\bs{m}_0^+, \bs{v}_0^+ }\right)
        \mathcal{N}_c\left({\bs{t}|\bs{m}_1^-, \bs{v}_1^- }\right)
    \, \mathrm{d}\bs{x}
\nonumber\\
= &
    \int_{\bs{x}}
    \delta(\bs{t}-\bs{Hx})
    \mathcal{N}_c\left({\bs{x}|\bs{m}_0^+,  \bs{v}_0^+}\right)
    \mathcal{N}_c\left({\bs{Hx}|\bs{m}_1^-, \bs{v}_1^-}\right)
    \, \mathrm{d}\bs{x}
\\
\propto &
    \int_{\bs{x}}
    \delta(\bs{t}-\bs{Hx})
    \mathcal{N}_c(\bs{x}|\bs{x}^+,\bs{Q}_x^+)
    \, \mathrm{d}\bs{x}
\label{eq:delta_intgral}
\end{align}
where
\begin{align}
\bs{Q}_x^+&=\left({\bs{H}^H\text{Diag}(\bs{1}\oslash \bs{v}_1^-)\bs{H}+\text{Diag}(\bs{1}\oslash \bs{v}_0^+)}\right)^{-1}\\
\bs{x}^+&=\bs{Q}_x^+ \left({\bs{H}^H\text{Diag}(\bs{1}\oslash \bs{v}_1^-)\bs{m}_1^- + \bs{m}_0^+ \oslash \bs{v}_0^+}\right)
\end{align}
The integration in (\ref{eq:delta_intgral}) actually yields a Gaussian density, whose mean and covariance are $\bs{H}\bs{x}^+$ and $\bs{HQ}_x^+\bs{H}^H$, respectively. Given this, the projection result then becomes
\begin{align}
\hat{\bs{t}}^+&=\bs{H}\bs{x}^+\\
\bs{v}_t^+&=\text{diag}(\bs{HQ}_x^+\bs{H}^H)
\end{align}
and we could further compute the extrinsic message
\begin{align}
\mathcal{N}_c(\bs{t}|\bs{m}_1^+, \bs{v}_1^+)
\propto &
\frac{\mathcal{N}_c\left({\bs{t}|\hat{\bs{t}}^+, \bs{v}_t^+}\right)}{\mathcal{N}_c(\bs{t}|\bs{m}_1^-, \bs{v}_1^-)}
\end{align}
where
\begin{align}
\bs{v}_1^+&=\bs{1}\oslash \left({\bs{1} \oslash \bs{v}_t^+ - \bs{1}\oslash \bs{v}_1^-}\right)\\
\bs{m}_1^+&=\bs{v}_1^+ \odot \left({\hat{\bs{t}}^+ \oslash \bs{v}_t^+ - \bs{m}_1^-\oslash \bs{v}_1^-}\right)
\end{align}

We continue with the factor node $p(\bs{y}|\bs{t})$, i.e., Module B,
\begin{align}
\mathcal{N}_c\left({\bs{y}|\hat{\bs{y}}^+, \bs{v}_y^+}\right)
%\nonumber \\
%= &
%\text{Proj}_{\bs{y}}
%\left[
%    \mathcal{N}_c \left({\bs{y}|\bs{m}_2^-, \bs{v}_2^-}\right)
%    \int_{\bs{t}}
%    p(\bs{y}|\bs{t})
%    \mathcal{N}_c\left({\bs{t}|\bs{m}_1^+, \bs{v}_1^+}\right)
%    \, \mathrm{d} \bs{t}
%\right]
%\\
= &
\text{Proj}_{\bs{y}}
\left[
    \mathcal{N}_c \left({\bs{y}|\bs{m}_2^-, \bs{v}_2^-}\right)
    \mathcal{P}(\bs{y})
\right]
\end{align}
where
$
\mathcal{P}(\bs{y})
\triangleq \int_{\bs{t}}
    p(\bs{y}|\bs{t})
    \mathcal{N}_c\left({\bs{t}|\bs{m}_1^+, \bs{v}_1^+}\right)
    \, \mathrm{d} \bs{t}
$,
and the result is
\begin{align}
\hat{\bs{y}}^+
& =
    \mathbb{E}
    \left[
        \bs{y} \,|\, \bs{m}_2^-, \bs{v}_2^-, \mathcal{P}(\cdot)
    \right]
\\
\bs{v}_y^+
& =
    \mathrm{Var}
    \left[
        \bs{y} \,|\, \bs{m}_2^-, \bs{v}_2^-, \mathcal{P}(\cdot)
    \right]
\end{align}
So, the extrinsic message can now be computed as
\begin{align}
\mathcal{N}_c(\bs{y}|\bs{m}_2^+, \bs{v}_2^+)
&\propto
\frac{\mathcal{N}_c\left({\bs{y}|\hat{\bs{y}}^+, \bs{v}_y^+}\right)}{\mathcal{N}_c(\bs{y}|\bs{m}_2^-, \bs{v}_2^-)}
\\
\bs{v}_2^+
&=
\bs{1}\oslash \left({\bs{1} \oslash \bs{v}_y^+ - \bs{1}\oslash \bs{v}_2^-}\right)
\\
\bs{m}_2^+
&=
\bs{v}_2^+ \odot \left({\hat{\bs{y}}^+ \oslash \bs{v}_y^+ - \bs{m}_2^-\oslash \bs{v}_2^-}\right)
\end{align}
So far, we have completed a back and a forward passing. Connecting the two, we attain an iteration of the algorithm.
%
%\section{density-to-R.V.}
%If $f_{\underline{\bs{w}}}(\bs{w})$ is the pdf of a random vector $\underline{\bs{w}}$ and $g: \mathbb{R}^m\rightarrow \mathbb{R}^n$, then $\underline{\bs{v}}=\bs{H}\underline{\bs{w}}$ is equivalent to
%\begin{align}
%f_{\underline{\bs{v}}}(\bs{v})=\int \delta(\bs{v}-\bs{Hw})f_{\underline{\bs{w}}}(\bs{w})\, \mathrm{d}\bs{w}
%\end{align}
%\proof
%If $\underline{\bs{v}}=g(\underline{\bs{w}})$, then
%\begin{align}
%\text{Pr}(\underline{\bs{v}}<\bs{v})=\text{Pr}(g(\underline{\bs{w}})\leq \bs{v})=\int \prod\limits_{i=1}^n u((\bs{v}-\bs{Hw})_i) f_{\underline{\bs{w}}}(\bs{w})\, \mathrm{d}{\bs{w}}
%\end{align}
%where $u(\cdot )$ is unit step function. Thus
%\begin{align}
%\text{Pr}(\underline{\bs{v}}<\bs{v})=\frac{\, \mathrm{d} }{\, \mathrm{d}\bs{v}}\int \prod\limits_{i=1}^n u((\bs{v}-\bs{Hw})_i) f_{\underline{\bs{w}}}(\bs{w})\, \mathrm{d}{\bs{w}}=\int \delta(\bs{v}-\bs{Hw})f_{\underline{\bs{w}}}(\bs{w})\, \mathrm{d}\bs{w}
%\end{align}
%On the country, if $f_{\underline{\bs{v}}}(\bs{v})=\int \delta(\bs{v}-\bs{Hw})f_{\underline{\bs{w}}}(\bs{w})\, \mathrm{d}\bs{w}$ holds, obviously, there is $\underline{\bs{v}}=\bs{H}\underline{\bs{w}}$ based on the properties of delta function.

%%%%%%%%%%%%%%%%%%%%%%%%%%%%%%%%%%%%%%%%%
\section{Particular Cases}

\subsection{Finite-Resolution Case}
In this case we are interested in the case where the observation $\bs{y}$ is acquired through a complex-valued quantizer $Q_c(\cdot)$. Specifically, each complex-valued quantizer $Q_c(\cdot)$ consists of two real-valued $B$-bit quantizers $Q(\cdot)$, which is defined as
\begin{align}
y_a & =
Q_c(y_a)
\triangleq
Q(\Re[y_a]) + \mathbb{J} Q(\Im[y_a]).
\end{align}
Hence, the resulting quantized signal $\bs{y}$ is given by
\begin{align}
\bs{y}
& =
Q_c(\bs{t} + \bs{w})
\end{align}
where $\bs{w} \sim \mathcal{N}_c(\bs{0}, \sigma_w^2 \bs{I})$ represents the AWGN. Let $\mathrm{up} (\cdot)$ and $\mathrm{low} (\cdot)$ denote, respectively, the upper and lower boundaries of the quantization interval in which the received value $y_a$ falls, then the transition probability $p( y_a | z_a)$ of interest can be expressed as
\begin{align}
&
p( y_a | t_a)
=
\nonumber\\
&
\left[
    \Phi
    \left(
        \frac{
            \mathrm{up} (\Re[y_a]) - \Re[t_a]
        }{
            \sigma_w/ \sqrt{2}
        }
    \right)
    -
    \Phi
    \left(
        \frac{
            \mathrm{low} (\Re[y_a]) - \Re[t_a]
        }{
            \sigma_w/ \sqrt{2}
        }
    \right)
\right]
\cdot
    \nonumber\\
&
\left[
    \Phi
    \left(
        \frac{
            \mathrm{up} (\Im[y_a]) - \Im[t_a]
        }{
            \sigma_w/ \sqrt{2}
        }
    \right)
    -
    \Phi
    \left(
        \frac{
            \mathrm{low} (\Im[y_a]) - \Im[t_a]
        }{
            \sigma_w/ \sqrt{2}
        }
    \right)
\right]
\end{align}
with $\Phi(\cdot)$ being the standard Gaussian CDF.
To see how these boundaries look like, take uniform quantization as an example and consider a step size of $\triangle$, the quantization output is then given as below (for ease of notation, we abuse the notation $y_a$ to denote $\Re[y_a]$ and $\Im[y_a]$)
\begin{align}
y_a & \in
\left\{
    (-1/2 + b) \triangleq
    ;\;
    b = -2^{B-1}+1, \ldots, 2^{B-1}
\right\}
\end{align}
with the lower and upper boundaries being
\begin{align}
\mathrm{low}(y_a)
& =
\left\{
\begin{array}{ll}
  y_a - \triangle/2
    & \mathrm{if} \; y_a \geq - (2^{B-1} -1) \triangle,  \\
  -\infty
    & \mathrm{otherwise}.
\end{array}
\right.
\nonumber \\
\mathrm{up}(y_a)
& =
\left\{
\begin{array}{ll}
  y_a + \triangle/2
    & \mathrm{if} \; y_a \leq  (2^{B-1} -1) \triangle,  \\
  +\infty
    & \mathrm{otherwise}.
\end{array}
\right.
\end{align}
Under such a uniform quantization setting, the two equation

$
\mathcal{L}(\bs{t}) \triangleq
\int_{\bs{y}} p(\bs{y}|\bs{t}) \mathcal{N}_c(\bs{y}|\bs{m}_2^-, \bs{v}_2^-)    \,\mathrm{d}\bs{y}
$,

$
\mathcal{P}(\bs{y})
\triangleq \int_{\bs{t}}
    p(\bs{y}|\bs{t})
    \mathcal{N}_c\left({\bs{t}|\bs{m}_1^+, \bs{v}_1^+}\right)
    \, \mathrm{d} \bs{t}
$,

\subsection{Infinite-Resolution Case}
AWGN case

\begin{align}
&
\mathcal{N}_c\left({\bs{t}|\hat{\bs{t}}^-, \bs{v}_t^- }\right)
\nonumber\\
=&
\text{Proj}_{\bs{t}}
\left[
    \mathcal{N}_c(\bs{t}|\bs{m}_1^+, \bs{v}_1^+)
    \int_{\bs{y}}
        p(\bs{y}|\bs{t})
        \mathcal{N}_c(\bs{y}|\bs{m}_2^-, \bs{v}_2^-)
    \,\mathrm{d}\bs{y}
\right]
\\
\propto &
\mathcal{N}_c(\bs{t}|\bs{m}_1^+, \bs{v}_1^+)\mathcal{N}_c\left({\bs{t}|\bs{m}_2^-, \sigma_1^2\bs{1}+\bs{v}_2^- }\right)
\end{align}
After using the Gaussian reproduction property, it becomes
\begin{align}
\bs{v}_t^-&=\bs{1} \oslash \left({\bs{1}\oslash \bs{v}_1^+ + \bs{1}\oslash (\sigma_1^2\bs{1}+\bs{v}_2^-)}\right)\\
\hat{\bs{t}}^-&=\bs{v}_t^- \odot \left({ \bs{m}_1^+\oslash \bs{v}_1^+ +\bs{m}_2^- \oslash (\sigma_1^2\bs{1}+\bs{v}_2^-)}\right)
\end{align}

******************

Given $p(\bs{y}|\bs{t}) = \mathcal{N}_c\left({\bs{y}|\bs{t}^-, \sigma_1^2 \bs{1}}\right)$, the integral above becomes
\begin{eqnarray}
&&
\int_{\bs{t}} p(\bs{y}|\bs{t})
\mathcal{N}_c\left({\bs{t}|\bs{m}_1^+,\bs{v}_1^+}\right)
\mathcal{N}_c\left({\bs{y}|\bs{m}_2^-,\bs{v}_2^-}\right)\, \mathrm{d}\bs{t}
\nonumber \\
& \propto &
\mathcal{N}_c\left({\bs{y}|\bs{m}_2^-, \bs{v}_2^-}\right)
\mathcal{N}_c\left({\bs{y}|\bs{m}_1^+, \bs{v}_1^++\sigma_1^2\bs{1}}\right)
\\
\bs{v}_y^+
& = &
\bs{1}\oslash \left({\bs{1} \oslash \bs{v}_2^- +\bs{1}\oslash (\bs{v}_1^++\sigma_1^2\bs{1})}\right)
\\
\hat{\bs{y}}^+
& = &
\bs{v}_y^+\odot \left({\bs{r}_1^+ \oslash (\bs{v}_1^++\sigma_1^2\bs{1})+\bs{m}_2^-\oslash \bs{v}_2^-}\right)
\end{eqnarray}

\end{appendices}

\bibliographystyle{IEEEtran}%
\bibliography{bib_zhc}%IEEEabrv

%\end{spacing} % 结束间距
\end{document}